\documentstyle[psfig]{l-aa}     

\def \B {\mbox{B}}
\def \Bj {\mbox{B}_J}

\def \Rf {\mbox{R}_F}

\def \rms {{\it rms}}
\def \mic {~{$\mu$m}}
\def \etal {{et~al.\ }}

\begin{document}
 
   \thesaurus{3(13.09.1;11.05.2;11.19.3;11.09.2;11.06.2)}
   \title{Galaxy evolution at low redshift? II. Number counts and optical identifications of faint IRAS sources}
   \author{E.~Bertin\inst{1,2}, M.~Dennefeld\inst{1,3}, and M.~Moshir\inst{4}}
   \offprints{E.~Bertin {\it (bertin@iap.fr)}}
   \institute{Institut d'Astrophysique de Paris, 98bis Boulevard Arago, 75014 Paris, France
	\and  Sterrewacht Leiden, PO Box 9513, 2300 RA, Leiden, The Netherlands
	\and  Universit\'e Pierre \& Marie Curie, 4, place Jussieu, F-75005 Paris, France
        \and  Infrared Processing and Analysis Center, California Institute of Technology, Pasadena, CA 91125, USA}
   \date{Received 10.7.1996; accepted}
   \maketitle 
\markboth{E.~Bertin, M.~Dennefeld, and M.~Moshir: faint IRAS sources}{}
   \begin{abstract}
%
   We analyse a Far InfraRed (FIR) catalogue of galaxies at 60\mic\ with a flux limit of $\approx 110$~mJy, extracted
   from a deep subsample of the IRAS Faint Source Survey. Monte-Carlo simulations and optical identification
   statistics are used to put constraints on the evolution of galaxies in the FIR.
   We find evidence for strong evolution of IRAS galaxies, in luminosity
   $\propto (1+z)^{3.2\> \pm 0.2\pm 0.3}$ or density $\propto (1+z)^{6.0\> \pm 0.5\pm 0.7}$
   for $f_{\nu}(60\mu{\rm m})> 150$~mJy,
   in agreement with previous claims.
   An excess of rather red faint optical counterparts with $\B>18$, consistent with the above evolution, is detected.
   We interpret these results as strong evolution at recent times among the starburst (or dusty AGN)
   population of merging/interacting galaxies. Most of these objects at moderate redshifts may pass unnoticed among
   the population of massive spirals in broad-band optical surveys, because of large amounts of dust extinction in
   their central regions. A possible link with the strong evolution observed in the optical for blue sub-$M^*$ galaxies is discussed.

   \end{abstract}
%
%

\section{Introduction}
  This is the second paper in a series where we try photometricaly to quantify galaxy evolution at low redshifts ($z<0.3$),
  from reasonably large and homogeneous samples. We have shown in Bertin \& Dennefeld (\cite{bertin:dennefeld}, paper I)
  that optical galaxy counts were consistent with little evolution down to $\Bj= 21$, corresponding
  to $z\approx 0.2$ for $M^*$ galaxies. In this paper, we investigate the case of the far-infrared domain.

  The Far-InfraRed (FIR) is very sensitive to ongoing star-formation activity in
  galaxies. Warm dust, essentially heated by the blue/UV continuum of young OB stars, makes a large contribution
  to the total FIR output of late-type objects. At the same time, a significant heating by an Active Galactic
  Nucleus (AGN) is possible in the most luminous FIR-emitting objects, although the respective contributions
  of star-formation and AGNs are not yet well determined.
  In this context, a statistical analysis of a deep and homogeneous FIR-selected galaxy sample
  should easily allow one to trace evolution of star-formation and/or AGN activity with redshift.

  The IRAS (InfraRed Astronomical Satellite) survey, with its four infrared channels at 12, 25, 60 and 100\mic\ 
  provides a unique tool for such studies,
  despite its moderate sensitivity. Based on the Point Source Catalog (PSC) and the more recent Faint Source Catalog (FSC)
  extracted from the IRAS data, several workers have investigated the evolution hypothesis on subsamples reaching various depths.
  The situation is still somewhat unclear, although evolution seems to be detected at 60\mic\ in most cases.
  Using PSC data, with solid identifications and radial velocities, Saunders
  \etal (\cite{saunders:ala}) claimed a very strong evolution effect in their sample, while Fisher \etal (\cite{fisher:al})
  found theirs compatible with no evolution (the latter has however a somewhat lower median redshift).
  Using the deeper FSC data, Lonsdale \etal (\cite{lonsdale:al}) found evidence for strong evolution from
  source number counts. Oliver \etal (\cite{oliver:alb}) reported also evolution in their spectroscopic subsample,
  though at a slightly milder rate.
  Two deeper studies (Hacking \& Houck \cite{hacking:houck} and Gregorich \etal \cite{gregorich:al}) made on much smaller
  specific areas (a few tens of square degrees) suggest a large excess of faint detections. 

  In order to clarify the problem and to complete these studies by providing the ``missing link'' between extensive, shallow
  catalogs ($f_{\nu}(60\mu{\rm m})> 0.2$~Jy) and small, very deep studies ($f_{\nu}(60\mu{\rm m})\la 100$~mJy),
  we have undertaken an identification program of IRAS sources at 60\mic\ on some selected areas in the sky,
  focusing ourselves especially on the faintest flux domain achievable with the Faint Source Survey (FSS)
  ``plates'':
  $f_{\nu}(60\mu{\rm m}) \approx 120$~mJy with a signal-to-noise ratio of 4.
  Our purpose is twofold: (1) to estimate the reliability of the IRAS survey at its very faint end,
  and (2), to
  characterize the population of IRAS galaxies in the 100 mJy domain (where evolution is expected to begin
  to show up in models of FIR galaxy counts).

  At this low S/N level, the data cannot be interpreted in a straightforward way: completeness, reliability and
  measurement biases must be quantified accurately. We rely here on both Monte-Carlo simulations and
  optical identifications to estimate these effects.

  The paper is organized as follows. In \S\ref{chap:data} we describe the selection of the infrared ``real'' sample.
  \S\ref{chap:simul} is devoted to the description of our Monte-Carlo simulations of
  IRAS catalogs. These are then used in \S\ref{chap:infra} to interpret statistics on the infrared data
  (confusion noise, number counts). In \S\ref{chap:optid} we rely on optical identifications to put further
  constraints on the nature of faint sources.
  We summarize our results in \S\ref{chap:comp} and compare them to those of other studies.
  In \S\ref{chap:discu}, we finally discuss the implications in the framework of galaxy evolution.

\section{The data}
\label{chap:data}
  The data in this study come from the Very Faint Source Sample (VFSS).
  The VFSS was generated by one of us (MM) after the completion of
  the FSS. The data in the FSS were searched for
  regions where 1) The IR cirrus indicators NOISRAT and NOISCOR (Moshir
  et al. 1992) were small, 2) the spacecraft coverage was at least
  4 HCONs and 3) there was a minimum number of nearby galaxies.
  These conditions led to a list of FSS plates which were then processed
  through the normal processing pipe-line.
  Their average ``instrumental'' noise is 25~mJy instead of about 40~mJy for more typical FSS plates.
  Since with these conditions, the reliability of the sample had been
  verified to be high through spot checks, the extraction thresholds for
  the normal FSS pipeline (which were set conservatively) were reduced
  (from S/N of 3 to 2.5) in order to increase the completeness
  of the VFSS down to  ~110 mJy at 60 \mic\ (and similar increases in
  completeness in the other 3 bands).
The resulting VFSS database covers ~400
square degrees in 4 separate contiguous regions (Table \ref{tab:vfsspos}) and contains
approximately 2,500 sources detected at 60\mic. The VFSS is about 2 times deeper than the FSC
and contains many unique sources since they can not be found in
any other IRAS catalogue or database\footnote{Although the field studied by Hacking \& Houck (1987)
has one of the highest spacecraft coverages and thus passes condition
2, it fails to pass condition 1 for cirrus contamination and was
automatically excluded.}.

\begin{table}[htbp]
\caption{IRAS fields\label{tab:vfsspos}}
\begin{tabular}{lrlrlr}
     \hline
   Field & \multicolumn{2}{l}{$\alpha$ (2000.0)} &
    \multicolumn{2}{l}{$\delta$ (2000.0)} & Area \\
   \cline{2-5}
   & h & m & \degr & \arcmin & (sq. degrees)\\
     \hline
     \hline
   N1   & 09 & 20 & +78 & 20 & 80.8\\
   N2   & 15 & 25 & +55 & 00 & 113.5\\
   N3   & 17 & 50 & +47 & 30 & 83.2\\
   S    & 03 & 50 & -47 & 00 & 133.7\\
     \hline
\end{tabular}
\end{table}

  Although we have only upper limits at 12, 25 and 100\mic\ for the great majority of the 60\mic\ detections, a few stars lie
  in the sample. We removed them by requesting infrared sources to have $f_{\nu}(25\mu{\rm m})<f_{\nu}(60\mu{\rm m})$.
  This simple criterion proves to filter efficiently most galactic objects, but might also discard a few Seyfert galaxies
  (e.g. Dennefeld \& Veron \cite{dennefeld:veroncetty}). However, as only 2\% of our detections do not pass the
  filter criterion (all identified as bright stellar sources in our optical subsample), we consider this effect to be negligible.

\section{The ``Monte-Carlo'' 60\mic\ sample}
\label{chap:simul}
  About 2/3 of our detections at 60\mic\ have a S/N lower than 5, i.e. a photometric accuracy
  worse than 20\%. At these levels, one expects
  the data to be seriously affected by completeness and reliability problems.
  Important photometric errors (if they are not purely fractional) do significantly bias number counts
  (Eddington 1913), as well as many other statistics that can be drawn from a flux limited survey.
  Murdoch \etal (\cite{murdoch:al}) have tabulated the Eddington bias, in the typical case of an underlying
  Euclidean slope and Gaussian flux errors, down to a ``safe'' $5\sigma$ level.
  For detections lying below this level, the simple analytical correction used by
  Oliver \etal (\cite{oliver:alb}) from the Murdoch \etal calculations, cannot be applied.
  The only escape is to proceed through Monte-Carlo simulations. This can be easily done here as both the
  instrumental and total noise levels are nearly constant over the surveyed zone. We have therefore
  created mock infrared images mimicking FSS plates, on which was run a source extraction algorithm similar to the
  original one used for the FSC. Statistics drawn from the simulation catalogs could then be directly compared with
  those from the real observations, with the hope of discriminating against different models of the
  IRAS galaxy population.

  The Monte-Carlo approach has some limitations at these low S/N levels,
  mainly because assumptions have to be made about
  the noise distribution. For this reason, we included only two components in the simulations:
  ``instrumental noise'', which we considered to be Gaussian, and galaxies.
  Given the very high
  coverage in our fields, the data conditioning involved in the production
  of FSS plates (including de-glitching, median filtering and trimmed average) should have removed all
  significantly non-Gaussian features from the instrumental noise distribution. Filaments from infrared cirrus (Low \etal \cite{low:al}) are more of a threat,
  and can easily create false detections by themselves. In what follows, we will keep in mind
  the possible existence of such noise components. But one can remark that the effect of transient noise sources like
  cirrus peaks and glitches, whose distributions are strongly skewed toward positive values,
  is essentially to lower the reliability of detections.

\subsection{The galaxy population}  
  We modeled the flux distribution of infrared galaxies assuming a constant
  comoving number density,
  and a Saunders \etal (\cite{saunders:ala}) Luminosity Function (LF):
  \begin{equation}
  \label{eq:lf}
  \phi(L_{60}) = C\left(\frac{L}{L^*}\right)^{1-\alpha}
  \exp\left[ -\frac{1}{2\sigma^2}\log^2_{10}\left( 1 + \frac{L}{L^*} \right) \right],
  \end{equation}
  where $L_{60} = \nu L_{\nu}$ at 60\mic, $\alpha = 1.09$,
  $\sigma = 0.724$ and $L^* = 2.95\,10^8\;h^{-2} L_{\odot}$. These best-fit parameters depend slightly on the value of $H_0$,
  which was conveniently set to 60\,${\rm km.s}^{-1}.{\rm Mpc}^{-1}$ in our simulation, close to the value of
  66\,${\rm km.s}^{-1}.{\rm Mpc}^{-1}$ adopted by Saunders et al.. $\phi(L_{60})$ was normalized so as to
  fit the QDOT galaxy counts (Rowan-Robinson \etal \cite{rowan:alb})
  over the domain $1<f_{\nu}(60\mu{\rm m})<2$~Jy. It
  therefore assumes that evolutionary effects are negligible at this flux level. We adopted the
  luminosity-dependent fit of Hacking \& Houck (\cite{hacking:houck}) to the Soifer~\etal (\cite{soifer:al})
  Bright Galaxy Sample, for the distribution of spectral indices.
  Pure luminosity evolution was introduced in some of the realizations by making
  the $L^*$ parameter evolve as $(1+z)^Q$. This simple form is not introduced here with strong physical justification,
  but only to provide a convenient estimator for the evolutionary rate of the galaxy population in the infrared.
  Luminosity evolution was preferred to density evolution as it is more convenient to simulate through Monte-Carlo methods.
  At the depth reached by our data, significant evolution can only be probed for FIR-luminous galaxies. As for
  these galaxies, the LF behaves much like a power-law (Eq. \ref{eq:lf}), very similar results are obtained for pure density evolution.
  We gave to the ``cool'' and ``warm'' components (Saunders \etal \cite{saunders:ala}) the same
  rate of evolution for simplicity. In fact the ``warm'' (starburst) component proves to be the essential contributor
  to the part of the LF where evolution can be probed here.

  The spatial distribution of galaxies was simulated using a three-dimensional log-normal density field
  (Coles \& Jones \cite{coles:jones}), and the Saunders \etal (\cite{saunders:alb}) autocorrelation function
  truncated at 20~Mpc and kept constant in spatial coordinates. The log-normal model gives
  a honest fit to the count probabilities in the 1.2~Jy sample (Bouchet \etal \cite{bouchet:al}) and,
  although it may
  be unable to reproduce very high density regions of features like sheets and filaments,
  it provides an interesting alternative to other simple ``toy universe'' construction methods
  (e.g. Soneira \& Peebles \cite{soneira:peebles}, Benn \& Wall 1995).

  The overall simulated galaxy sample was then placed in an Einstein-de~Sitter universe, within a rod
  $150~h^{-1}$~Mpc wide, extending up to $z=2$. This redshift limit was chosen here for
  practical purpose. It does not affect by more than a few \% counts with $f_{\nu}(60\mu{\rm m})> 10$~mJy
  for the highest evolution rates considered in this paper. Although (projected)
  clustering was obvious on simulation images as seen by eye, none of the statistical measurements discussed
  in this paper proved to be affected significantly if we switched to a simple
  Poissonian distribution. Nevertheless, all numbers presented here were derived from simulations {\em with} clustering.

\subsection{The images}
  For the sake of simplicity, we simulated directly IRAS images by projecting galaxies over the frame,
  without including the complex co-adding of individual scans.
  The simulated images were given the same pixel size as the original FSS plates: 0.5\arcmin. This generously
  samples the reconstructed IRAS ``point spread function'', thereby removing the need for oversampling when projecting
  the infrared sources before convolution. At the same stage, we added white noise to account for
  instrumental noise. Images were convolved by Fast-Fourier Transform with a 3.7\arcmin$\times$2.1\arcmin Gaussian beam
  (a reasonable fit to the 60\mic\ FSS templates at these ecliptic latitudes), with the minor axis (average
  ``in-scan'' direction) aligned with the y-axis. Finally, we suppressed the mean background
  and low spatial frequencies by applying the standard 60\mic\ zero-sum median filter of the FSS
  (Moshir \etal \cite{moshir:al}) along the y-axis of the image. In the FSS, this point-source
  filtering was normally done on individual IRAS scans, which implies some angular dispersion in the final images
  at high ecliptic latitude. Anyway, our simulated ``in-scan'' point-source profiles are in excellent agreement
  with those from the original plates.


\subsection{Source extraction}
  The mock catalogue was created from the simulated images by running a slightly modified version of
  the SExtractor (Bertin \& Arnouts \cite{bertin:arnouts}) photometry program,
  tuned to mimic closely the original FSC ``Extractor'' from IPAC (Moshir \etal \cite{moshir:al}).
  In particular, the standard ``clipped \rms'' estimate of background noise in SExtractor was replaced
  by the ``68\%-quantile'' from the FSC extractor, and the integrated flux of detected sources by
  the peak pixel intensity, as for the real data.

\section{Infrared statistics}
\label{chap:infra}
\subsection{Confusion noise}
\label{chap:fluctu}
  It is interesting to compare the average 68\%-quantile measured on simulations with those from the real flux-maps
  (Table \ref{tab:confnoise}). This parameter is particularly sensitive to the density of undetected sources:
  in quadrature, about 2/3 of the confusion noise is expected to come from sources in the 10-100~mJy flux range with
  the models considered here.
  The 68\%-quantile is fairly stable in our simulations and in the real measurements, and enables us to constrain the number
  of these faint galaxies. Model-dependent conclusions about the evolution rate can be drawn from it (there is of course
  a dependence on model normalisation itself, but number counts of bright sources should restrict this uncertainty to less
  than $\approx 10\%$).
  As Table \ref{tab:confnoise} shows, the fluctuations predicted by a model evolving as $L^*\propto(1+z)^Q$, with $Q\ga4$
  are too large (the uncertainty quoted here represents the \rms\ of the 68\%-quantile over the southern subset area, where it reaches its
  minimum value).
  The $Q=3$ and $Q=3.5$ models agree well with the data if there is no other contribution
  than pure instrumental noise affects in our FSS fields. Milder-evolution and no-evolution models may require additional
  noise, cirrus for example, to fit the data.

\begin{table}[htbp]
\caption{68\% quantile statistics in mJy (including 25 mJy \rms\ instrumental
noise).\label{tab:confnoise}}
\begin{tabular}{lcccc}
     \hline
     VFSS data & no evol. model & $Q=3$ & $Q=3.5$ & $Q=4$ \\
     \hline
     \hline
     $30.2 \pm 1.2^a$ & 28.1 & 30.1 & 30.7 & 32.8 \\
     \hline
\end{tabular}\\
{$^a$ \small standard deviation (see text).}
\end{table}

  Some very faint cirrus can unambiguously be seen in the unfiltered 60\mic\ maps, which means they
  may somehow contribute to the background confusion noise. Gaultier \etal (\cite{gautier:al})
  have estimated the confusion noise at 60\mic\ due to cirrus to be about 7~mJy \rms,
  at high galactic latitude in the FSC. This might be taken as a typical value for our catalog, as our fields
  are located at only $|b|\approx 35$, but are amongst the ``cleanest'' of the infrared sky concerning cirrus emission,
  and are comparable to higher galactic latitude zones. Because the infrared surface
  brightness distribution of cirrus has generally much larger wings than a Gaussian, one should
  expect their contribution to the total 68\%-quantile to be less than 7~mJy in quadrature.
  However, given the large uncertainties in the real importance of cirrus emission in our fields,
  we do not discard the possibility of significant cirrus contamination of both background
  and individual source measurements.

  In any case, this fluctuation analysis is consistent with previous estimates that used the integrated infrared background
  (Oliver \etal \cite{oliver:ala}), although it is slightly more stringent for the $Q=4$, $z_f=2$ model which is only marginally
  compatible with our data.

\subsection{Number counts}
\begin{figure}[htbp]
  \centerline{\psfig{figure=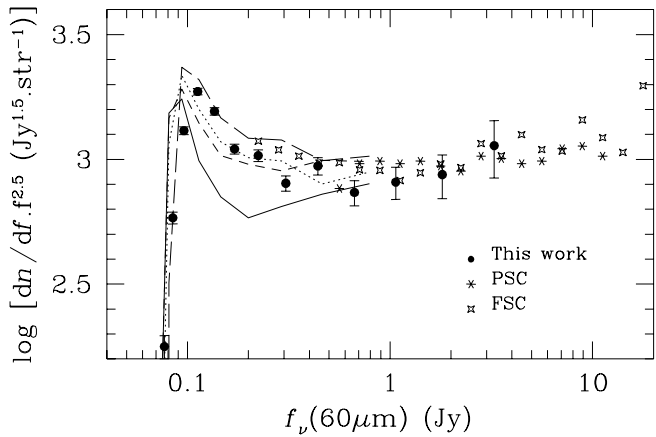,width=8.8cm}}
  \caption[]{
            Raw number counts compared with models at 60\mic. The PSC/QDOT (Rowan-Robinson \etal \cite{rowan:alb})
            and FSC (Lonsdale \etal \cite{lonsdale:al}) counts have been added for comparison. The FSC counts are
            not corrected for the Eddington bias, and so their two faintest bins are probably overestimated.
            For clarity, the Monte-Carlo models were plotted only for fluxes below 1~Jy.
            {\em solid}: no-evolution + Gaussian noise (see text); {\em dashes}: evolution with $Q=3$; {\em dots}: $Q=3.5$;
            {\em long dashes}: $Q=4$. Error bars are Poissonian uncertainties. Monte-Carlo models become very
            noisy brightward of 1 Jy, and have not been drawn there for clarity. Note that the bin size is
            not constant.
  \label{fig:rawcounts}
            }
   \end{figure}

  Deriving (raw) infrared number counts is straightforward here, as we are dealing with a homogeneous set of data.
  Comparison with our Monte-Carlo samples is shown in Fig. \ref{fig:rawcounts}. The Eddington bias is clearly
  visible below $f_{\nu}(60\mu{\rm m})\approx 200$~mJy. The best fit value for $Q$ proves to depend on the flux range
  considered. For $f_{\nu}(60\mu{\rm m})> 150$~mJy, i.e., for sources above with S/N higher than 5, we find $Q=3.2\pm0.2$,
  which however raises to $Q=3.6\pm0.3$ if we consider only the 0.15-0.2~Jy range. In the more uncertain 0.11-0.15~Jy domain,
  we find $Q=3.9 \pm 0.2$. Accounting for a $\pm10$\% possible error in the model normalisation yields a $\approx \pm0.3$ additional
  uncertainty to the values of $Q$ quoted above. These numbers may indicate that the $(1+z)^Q$ dependence in luminosity
  is  inappropriate for describing the effect seen here. However, we shall remember that, as stressed earlier,
  assumptions about the noise distribution become very critical in the 0.11-0.15~Jy bin and will affect conclusions made with
  low S/N data.

  This brings the question of whether our fields represent a ``fair sample''
  of the Universe, i.e. is there a relative excess of faint or bright sources due to large scale-structures
  which could distort the statistics (see e.g. Lonsdale \& Hacking \cite{lonsdale:hacking})?
  Fortunately this point can be addressed using our bright optical galaxy counts in the same areas (paper I).
  Expressed with the same magnitude scale, they are similar to those done on much larger catalogs
  (APM, COSMOS) in the $17<\Bj<21$ range, where most of the optical counterparts to faint IRAS sources lie.
  A strong distortion of galaxy counts induced by the presence of large scale structures is thus excluded.

  An interesting question is: what happens to number counts if we add
  some Gaussian noise to the no-evolution model, in order to reach the observed intensity of fluctuations? The
  answer is: almost nothing, as can be seen in Fig. \ref{fig:rawcounts}. As expected, this test indicates
  that the no-evolution model could be saved only if the distribution of ``hidden'' noise components is
  extremely non-Gaussian, creating noise peaks without increasing too much the 68\% quantile.

\section{Optical identifications}
\label{chap:optid}
  In order to find out whether the faint source excess measured in the previous sections is real or contaminated by
  noise peaks, some identification work is necessary. Systematic studies conducted at higher IRAS flux limits
  (e.g. Wolstencroft \etal  \cite{wolstencroft:ala}, \cite{wolstencroft:alb},
  Sutherland \& Saunders \cite{sutherland:saunders}) have shown that the optical domain is well adapted
  to this task.

\subsection{The data}
  Thirty percent (6 Schmidt plates) of the VFSS were analysed using photographic data in the blue $\Bj$ and red $\Rf$ passbands.
  The sensitivity of deep survey plates is well matched to our IRAS study, as the VFSS error box at
  the faintest flux levels occupies roughly 1 square-arcminute on the sky, and contains already an average
  of $\approx 0.3$ galaxies brighter than $\Bj = 21$.
  Full details about the optical catalog
  can be found in Paper I; we summarize here its main features. This catalogue is quite small compared to large galaxy
  surveys like APM (e.g. Maddox \etal \cite{maddox:alb}) or COSMOS (e.g. Heydon-Dumbleton \etal \cite{heydon:al}),
  but is well calibrated (within $\approx 0.1$~mag, thanks to a high density of photometric standards).
  The completeness for galaxies is expected to be better than a conservative 90\% at least for $17 \le \Bj \le 20$.
  Brighter than this, there might be some loss due to inaccurate star-galaxy separation, and underestimation of fluxes 
  ($\approx 0.1$~mag) because of photographic saturation effects. Positions are accurate within a fraction of arc-second.
  This is sufficient for identifying faint IRAS sources, which have typical positional uncertainties of a fraction of
  arcminute. 

\subsection{The simulated optical sample}
  As for the infrared statistics presented in the previous sections,
  extreme caution should be taken in interpreting the results of optical identifications.
  We don't know a priori what fraction of {\em real} IRAS sources can be identified, especially given the large errors in
  $f_{\nu}(60\mu{\rm m})$ at our flux level.
  Here again, we addressed this question using a Monte-Carlo method, by simulating a sample of identified optical
  counterparts to the artificial sample described in \S\ref{chap:simul}.

\subsubsection{Optical luminosity}
\begin{figure}[htbp]
  \centerline{\psfig{figure=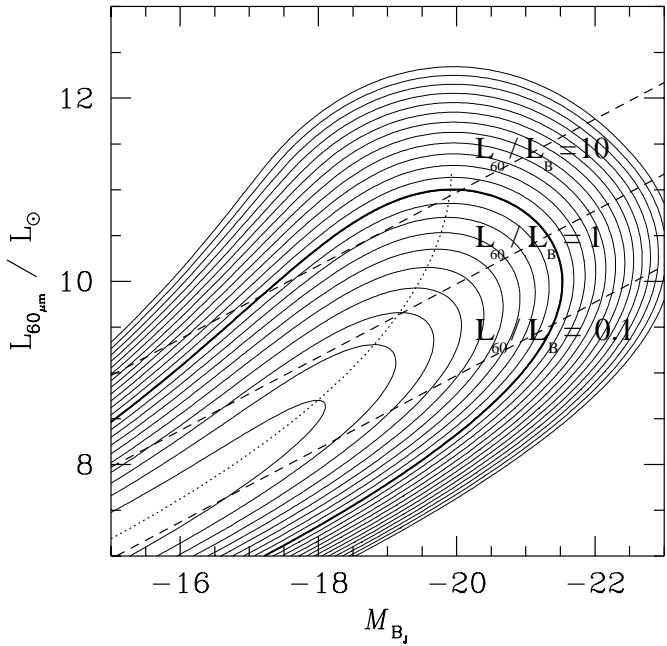,width=8.8cm}}
  \caption[]{
            Contour plot of the assumed bivariate 60\mic/$\Bj$ galaxy luminosity function, from the parameterization work of
            Saunders \etal (\cite{saunders:ala}). There is a factor 2 in density from one contour to another.
            The luminosity range analysed by Saunders \etal is approximately
            delineated by the heavier contour. The dotted curve represents Eq. (\ref{eq:bilf}).
            Dashes indicate lines of constant $L_{60}/L_B$.
  \label{fig:bilf}
            }
   \end{figure}

  We first assigned some random optical luminosity to each infrared ``Monte-Carlo'' galaxy with theoretical
  flux received on Earth $f_{\nu}(60\mu{\rm m})>10$~mJy, according to a suitable optical/60\mic\ distribution.
  This is an easier task than it may appear, as IRAS and bright optical galaxy catalogs do
  have an overlap large enough so that one can build a local, bivariate luminosity function. The most comprehensive
  study in that field was done by Saunders \etal (\cite{saunders:ala}), who proposed a Gaussian distribution of
  the blue absolute luminosity, with standard deviation $\sigma_M = 0.72$ and mean
  \begin{equation}
  \label{eq:bilf}
  \mu_M = -19.9 + 5\log h + 0.3\left(\log \frac{L_{60}}{L_{\odot}} - 11.24 - \log h^{-2}\right)^2.
  \end{equation}
  Note that the original expression contains a sign error.
  $M$ normally refers here
  to a $B^0_T$ magnitude, which corresponds almost to our $\Bj$ magnitude system (Paper I)
  as we drop the average correction for internal extinction $A_i = 0.21$ (Rowan-Robinson \cite{rowan:ala}).
  Although (\ref{eq:bilf}) is based on limited statistics and rather inaccurate optical magnitudes, we believe
  it does provide a globally correct description of the link between the 60\mic\ and blue luminosity functions
  because:
  1) relation (\ref{eq:bilf}) correctly reproduces both the $L_{FIR}/L_B$ distribution
  for ``normal'' luminosity sources (Rowan-Robinson \etal \cite{rowan:ala},
  Bothun \etal \cite{bothun:al}) and the ``saturation'' of optical luminosities observed in Ultraluminous
  InfraRed Galaxies (ULIRGs) (Smith \etal \cite{smith:al}, Soifer \etal \cite{soifer:al});
  2) integrating over all possible infrared luminosities leads to a blue luminosity function in excellent
  agreement with recent, local determinations\footnote{It is interesting to note that this leads to a ``high''
  normalisation of the {\em local} blue luminosity function, in agreement with our bright optical counts
  (Paper I).}
  (e.g. Ellis \etal 1996). The bivariate luminosity function based on (\ref{eq:lf}) and (\ref{eq:bilf}) is shown in
  Fig. \ref{fig:bilf}. We assume relation (\ref{eq:bilf}) to hold out to $z \approx 0.3$, corresponding to the
  most distant, luminous galaxies we might detect in numbers both in FIR and in the optical.
  This obviously supposes a negligible evolution of their dust content over the corresponding period of time. We
  shall tackle this point in \S\ref{chap:discu}.

\subsubsection{$k$-corrections}
\label{chap:kcorr}
  In order to give the right optical $k$-corrections to our simulated IRAS sources, one has to make some general assumption about their
  optical/UV spectral energy distribution (SED). The large majority of IRAS galaxies contained in
  FIR flux-limited samples are spirals (e.g. de Jong \etal \cite{dejong:ala}), with average morphological
  type Sbc (Bothun \etal \cite{bothun:al}, Sauvage \& Thuan \cite{sauvage:thuan}).
  Although their rest-frame optical colours span a considerable range, it is difficult to find an obvious trend
  with FIR luminosity (Leech \etal \cite{leech:ala}, \cite{leech:alb}, Hutchings \& Neff \cite{hutchings:neff}).
  An Sbc SED also proves to reproduce satisfactorily
  the observed optical continuum of 1.4-GHz sub-mJy starburst galaxies with $B<22$ (Benn \etal \cite{benn:al}). Given
  the tight correlation existing between radio-continuum and far-infrared galaxy luminosities
  (de Jong \etal \cite{dejong:alb}, Helou \etal \cite{helou:al}),
  $f_{\nu}({\rm 1.4 GHz}) \approx 10^{-2} f_{\nu}(60\mu{\rm m})$, these sources are representative of the faintest
  ones we might detect in our 60\mic\ sample. We have therefore adopted the Sbc SED from Pence (\cite{pence})
  to compute optical $k$-corrections through our $\Bj$ and $\Rf$ passbands, and the average Sbc
  $\langle\Bj - \Rf\rangle = 0.96$ typical rest-frame colour from Paper I. We will find in \S\ref{chap:colour} other arguments
  justifying this choice.

  We finally end up with a large number of artificial sources having in addition to their ``unbiased'' $f_{\nu}(60\mu{\rm m})$ fluxes,
  $\Bj$ and $\Rf$ magnitudes that could directly go through the same complex IR detection and optical identification
  processes, as real galaxies.

\subsection{Identification method}
  Because of the low resolution of IRAS images ($\approx 2'$ at 60\mic), the reliability of the identification technique
  is crucial for finding the right optical counterparts to faint infrared sources. This is even more critical here,
  because of the particularly low infrared S/N and optical faintness for most of the candidates. We have adopted the
  likelihood-ratio method, inherited from radio-source identification programs (de~Ruiter \etal \cite{deruiter:al},
  Prestage \& Peacock \cite{prestage:peacock}), which was successfully applied in previous IRAS studies
  (Wolstencroft \etal \cite{wolstencroft:ala},\cite{wolstencroft:alb}, Sutherland \& Saunders \cite{sutherland:saunders}).

\begin{figure}[htbp]
  \centerline{\hbox{\psfig{figure=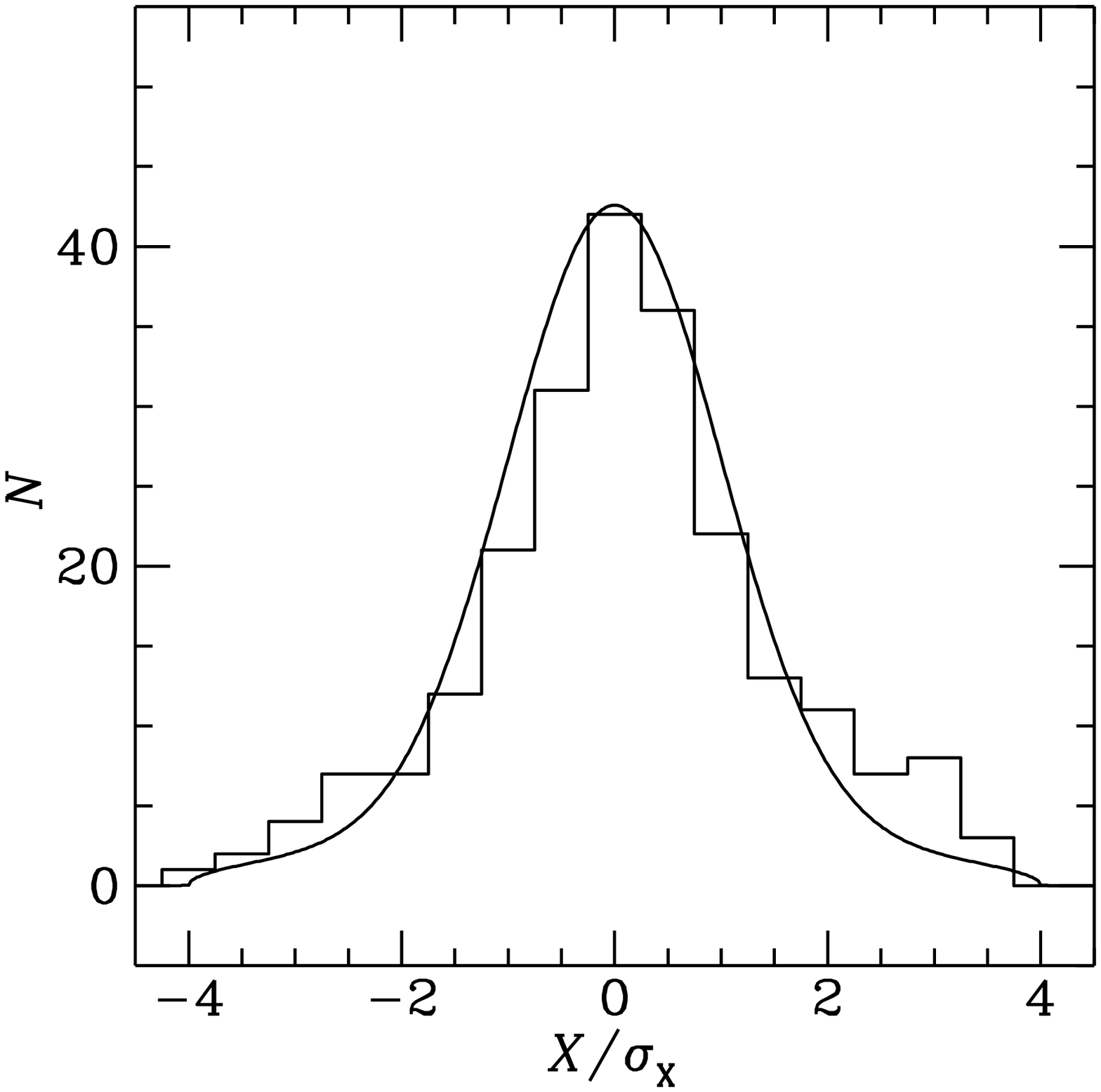,width=4.4cm}\psfig{figure=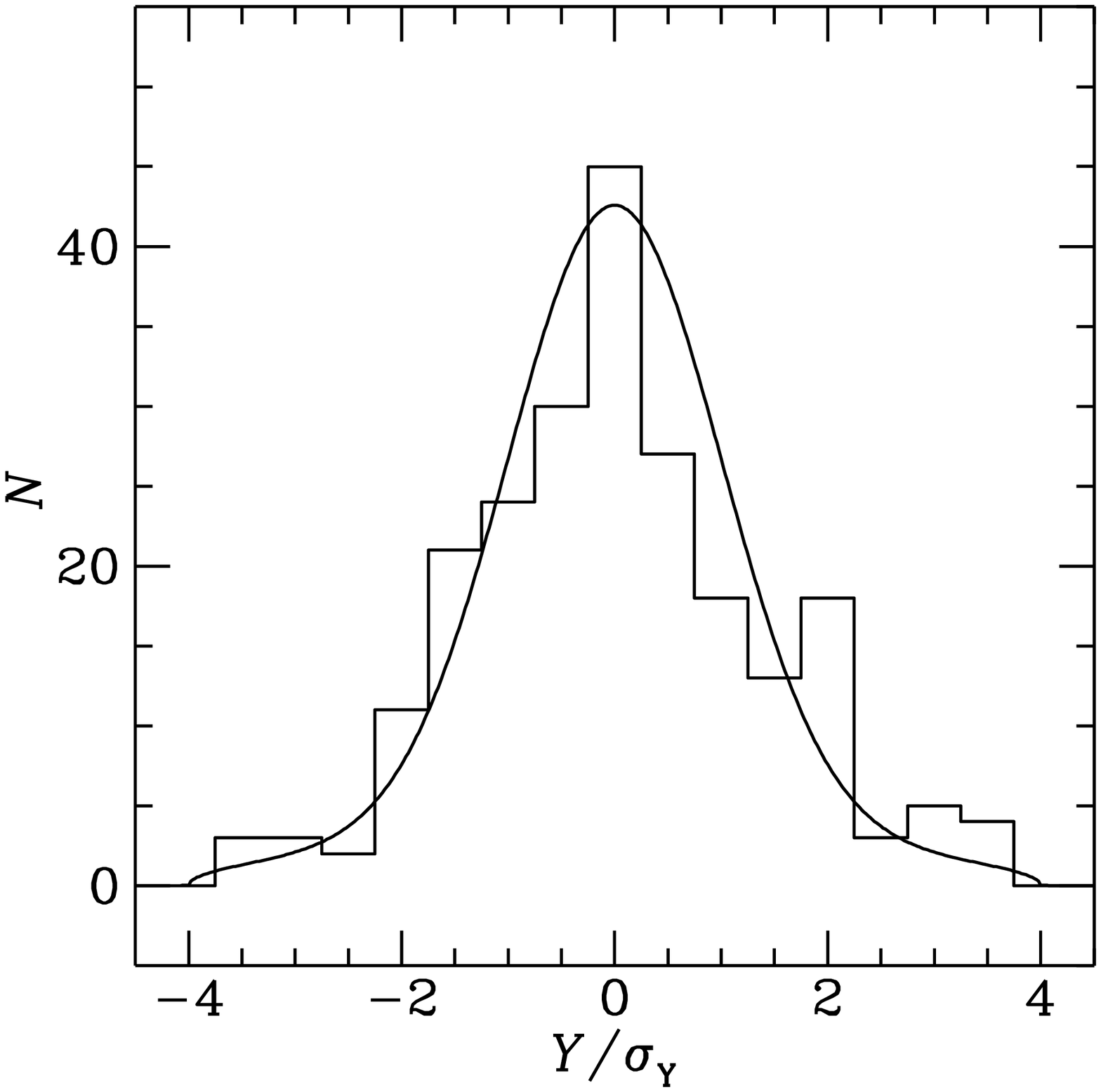,width=4.4cm}}}
  \caption[]{
            Distribution of the 227 galaxies with $\Bj<17$ and $r \le 4$ detected around infrared sources,
            along the major (left) and minor axes
            of the IRAS error ellipse (histogram). When several galaxies pass this criterion, the barycenter of the group is taken.
            The model (curve) assumes Gaussian positional errors and take into account contamination by ``chance'' counterparts
            ($\approx 6$\% at center of ellipse).
  \label{fig:eldistrib}
            }
  \end{figure}

  The positional uncertainty of an IRAS detection is essentially a function of the local reconstructed IRAS point spread function
  and the S/N. Assuming the error distribution is roughly Gaussian, this defines the traditional error ellipses around the
  estimated object position, described by the $\sigma_x$, $\sigma_y$ and $\theta$; respectively the \rms\ error along the major axis
  (cross-scan), the \rms\ error along the minor axis (in-scan), and position angle. Switching to the reduced coordinate:
  \begin{equation}
  r = \sqrt{\frac{X^2}{\sigma^2_X}+\frac{Y^2}{\sigma^2_Y}},
  \end{equation}
  we can express the probability for the true optical counterpart to lie between $r$ and $r+dr$ as
  \begin{equation}
  p_r(r)\,dr = Q\,e^{-r^2/2}\,r\,dr
  \label{eq:prid}
  \end{equation}
  where Q is the probability for the infrared detection to exist and to be detectable in the optical catalogue.
  In order to check that this modeling fits the data correctly, we derived maximum-likelihood
  estimates of $\sigma_x$, $\sigma_y$ and $\theta$ from optical identifications of the brightest infrared sources
  ($f_{\nu}$(60\mic)$> 0.2$~Jy). We were able to recover the catalogue values within 10-20\%.
  Of course, this does not mean that the modeling is {\em adequate}. There is indeed some evidence that the wings of
  the positional error distribution are larger than Gaussian ones (e.g. Sutherland \& Saunders
  \cite{sutherland:saunders}); we will quantify the consequences of this effect in the next section.
  As shown in Fig. \ref{fig:eldistrib}, the Gaussian model proves to fit correctly the distribution of optically bright ($\Bj<17$)
  counterparts in our catalogue. One can however already detect a slight excess ($\approx 5$\% of the total) of identifications
  at about $3\sigma$ along the cross-scan direction.

  Equation (\ref{eq:prid}) defines the probability density for a genuine identification to lie around the
  IRAS detection. We can compare it to the probability of encountering by chance an optical object of class $c$,
  brighter than magnitude $m$ between $r$ and $r+dr$:
  \begin{equation}
  p_f\,(r,m,c)\,dr = 2\pi \sigma_X \sigma_Y N_c(m)\,r\,dr,
  \label{eq:prnonid}
  \end{equation}
  where $N_c(m)$ is the projected density of optical objects with class $c$ brighter than $m$.
  This leads to the likelihood ratio definition from Wolstencroft \etal (\cite{wolstencroft:ala}):
  \begin{equation}
  L\,(r,m,c) = \frac{p_r(r, m, c)}{p_f(r,m,c)} = \frac{Q\,e^{-r^2/2}}{2\pi \sigma_X \sigma_Y N_c(m)}.
  \label{eq:lrdef}
  \end{equation}
  Sutherland \& Saunders (\cite{sutherland:saunders}) stressed that $L$ is independent of the number of
  optical candidates lying in the error ellipse. Some extra information can be gained from the presence
  or absence of these other candidates if one assumes that the true optical counterpart to the infrared source
  is unique. However, this hypothesis is dangerous for IRAS galaxies, because of the non-negligible fraction
  of multiple systems optically identified (e.g. Gallimore \& Keel \cite{gallimore:keel}). Besides, the artificial
  and real optical galaxy samples do not have the same density of ``background'' galaxies, as the first one
  comes from an infrared flux-limited sample: any estimator that combines information from the other optical detections
  is likely to behave differently in both cases. We therefore conformed to
  the ``standard'' likelihood ratio $L$ as an {\em individual} identification estimator for optical candidates.

  For every IRAS detection lying within the area of our optical catalog, likelihood ratios $L_i$ associated
  to every potential optical candidate were obtained in the following way. First, we estimated the $N_c(\Bj)$ separately
  for stars ($c=S$) and galaxies ($c=G$) from the number counts relative to the current plate. Then we defined a circular zone
  with a 3\arcmin\ radius around each IRAS position. This corresponds to a large confidence interval for $r$:
  at least $6\sigma$, or a diameter 3 times larger than the IRAS beam FWHM. Finally, the $L_i$ were computed
  using (\ref{eq:lrdef}) for all optical detections within the circular zone. The same procedure was applied to the artificial
  optical sample around detections in the simulated 60\mic\ images, but this time using $N_G(m) = {\rm dexp}\, 0.5(\Bj-15.0)$,
  which provides a good fit to our integrated galaxy number counts per square degree (paper I). The sizes of the
  ``artificial'' error ellipses were computed with formulae (II.F.5) and (II.F.6) in Moshir \etal (\cite{moshir:al}),
  assuming the number of scans to be $\langle N \rangle = 20$, and a background noise of 30.2 mJy.

\subsection{Identification rate}
  What {\em real} fraction of infrared sources possess one or more optical counterparts?
  Using the individual likelihood
  ratios $L_i$, we can define a global estimator ${\cal L}$ for each IRAS detection,
  from which we will derive an identification ratio. Here again, we need to be able to compare statistics
  done with different densities of contaminating sources. For this reasons, we have chosen
  \begin{equation}
  \label{eq:cali}
  {\cal L} \equiv {\rm sup}(L_i),
  \end{equation}
  though it might certainly be less effective for multiple sources than, e.g. ${\cal L} \equiv \sum_i L_i$.
  In the presence of contamination by ``chance'' optical objects, the observed probability for ${\cal L}$ [as defined by (\ref{eq:cali})]
  to exceed some threshold ${\cal L}_0$ for an infrared detection reads:
  \begin{equation}
  p\,({\cal L}>{\cal L}_0) = p_r({\cal L}>{\cal L}_0)\left( 1-p_f({\cal L}>{\cal L}_0)\right) + p_f({\cal L}>{\cal L}_0),
  \end{equation}
  where $p_r({\cal L}>{\cal L}_0)$ and $p_f({\cal L}>{\cal L}_0)$ are the probabilities to have
  ${\cal L}>{\cal L}_0$ because of real and fortuitous optical identifications, respectively. Thus
  \begin{equation}
  p_r = \frac{p - p_f}{1 - p_f}.
  \end{equation}
  $p_f$ can be measured by putting error ellipses at random positions in both real
  and artificial catalogs, which then enables us to compare the real identification rates $p_r$.

\begin{figure}[htbp]
  \centerline{\psfig{figure=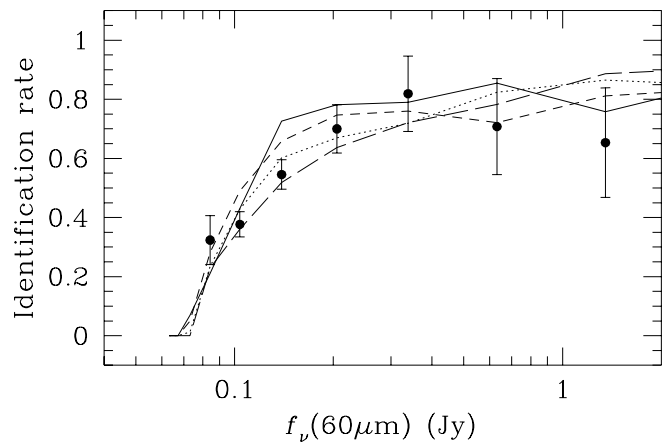,width=8.8cm}}
  \caption[]{
            Identification rate (corrected for the effects of spurious identifications) of IRAS sources with a
            likelihood threshold of 5. The lines are the Monte-Carlo models, as in Fig. \ref{fig:rawcounts}.
  \label{fig:idrate}
            }
   \end{figure}

  Figure \ref{fig:idrate} shows $p_r$ as a function of the 60\mic\ flux, for ${\cal L}_0 = 5$
  (which proves to be a good compromise
  between completeness and reliability). Although they do not enable us to choose between the models, these statistics
  seem to indicate that the faintest infrared sources cannot be attributed to a large number of unexpected ``noise peaks''.
  Another interesting result is the rather low identification rate predicted by simulations at high S/N: about 80\%.
  This in good agreement with the observations, and the $\approx 90$\% identification rates reported for the FSC
  (Sutherland \etal \cite{sutherland:al}, Wolstencroft \etal \cite{wolstencroft:alb}),
  once the differences in the sizes of error ellipses and the addition of chance identifications have been accounted for.
  An examination of the simulated FSS images reveals that the unidentified sources are produced by confusion with
  close neighbours in projection. Because of the large width of the IRAS beam,
  the barycenter of some detections can be shifted by a few pixels, i.e. $5 \sigma$ or more at high S/N, adding a
  a long tail to the Gaussian error distribution. For these detections, the measured likelihood ratio of the
  optical counterpart(s) is thus very close to zero. We conclude that this phenomenon might be responsible of
  most of the high S/N unidentified point-sources in the FSC. Note that cirrus sources do also constitute potential
  deflection factors.

\subsection{Optical colour distribution}
\label{chap:colour}
\begin{figure}[htbp]
  \centerline{\psfig{figure=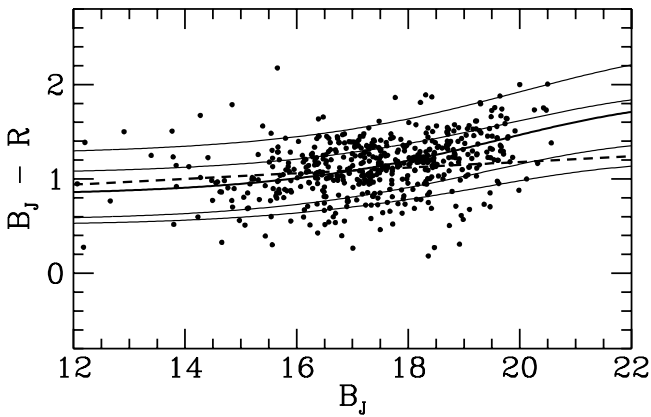,width=8.8cm}}
  \caption[]{
            Colour-magnitude diagram of all optical counterparts to IRAS sources with ${\cal L} > 10$, in the
            photographic $\Bj$ and $R$ bands (points). This high threshold in likelihood limits the proportion of fortuitous
            sources to about 10\%. Solid curves are the average colour tracks predicted by the
            non-evolving model described in the text, assuming different $k$-corrections.
            From top to bottom: E/S0, Sab, Sbc ({\em heavy line}\,), Scd and Sdm. For comparison, the dashed line indicates
            the average colour of our complete optical galaxy sample as a function of magnitude.
  \label{fig:colmag}
            }
   \end{figure}

  The colour distribution for all identifications with ${\cal L} > 10$ (roughly a 10\% contamination)
  is shown in Fig. \ref{fig:colmag}, together with the average colour tracks produced by the differential $k$-corrections
  computed with the Pence (\cite{pence}) SEDs. For clarity, only the no-evolution model is plotted. The $Q=3,4$ models yields
  tracks very similar to those of the no-evolution model, which is normal as Eq. (\ref{eq:bilf}) provides very little
  evolution in optical luminosity for luminous FIR sources.
  The dispersion in colours appears to be substantially smaller than for an optically limited sample, and as can be seen,
  the bulk of the IRAS population follows quite closely the Sbc track, supporting
  our choice in \S\ref{chap:kcorr} to adopt an Sbc type for the average optical $k$-correction.

\subsection{Magnitude distribution}
\begin{figure}[htbp]
  \centerline{\psfig{figure=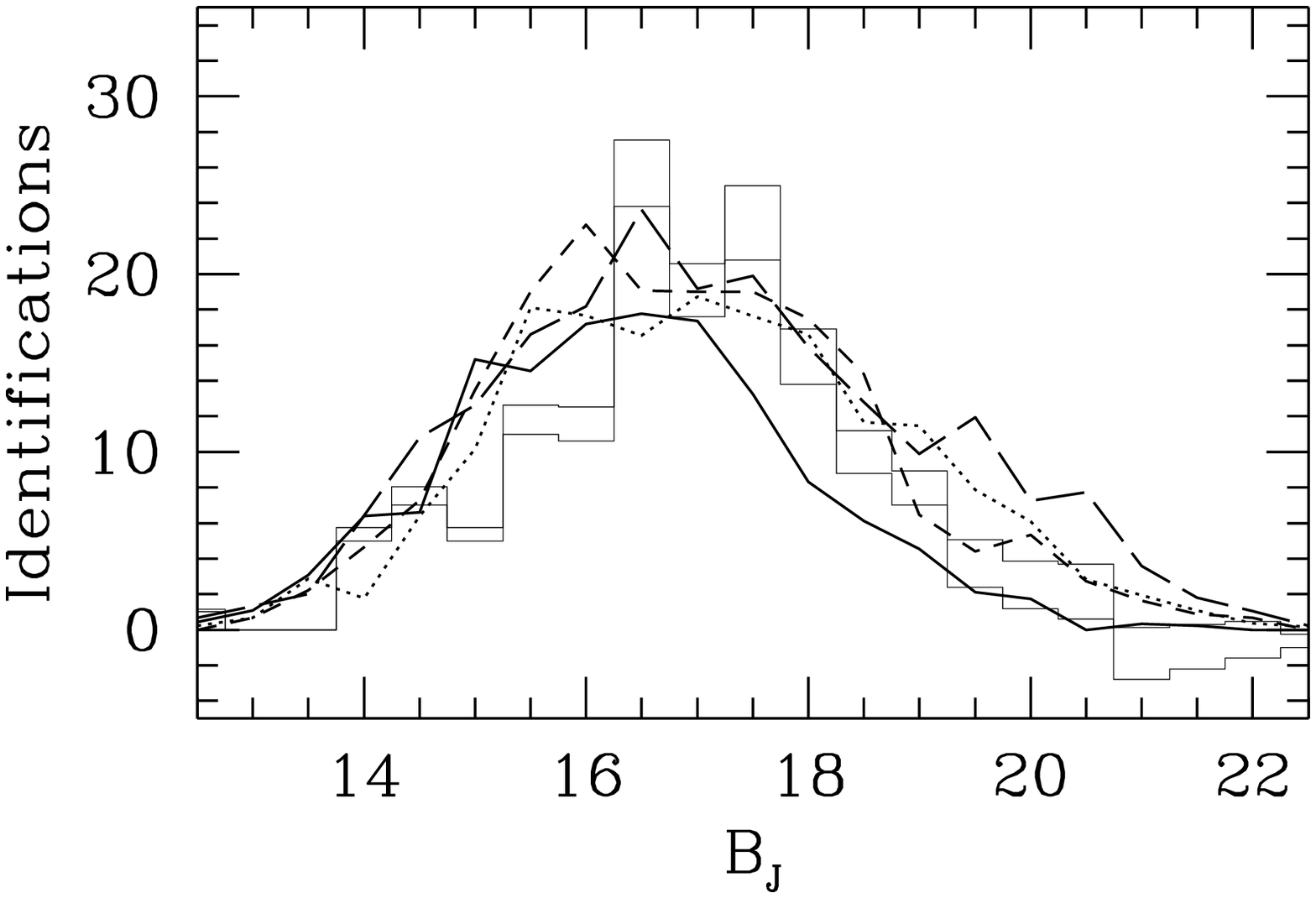,width=8.8cm}}
  \centerline{\psfig{figure=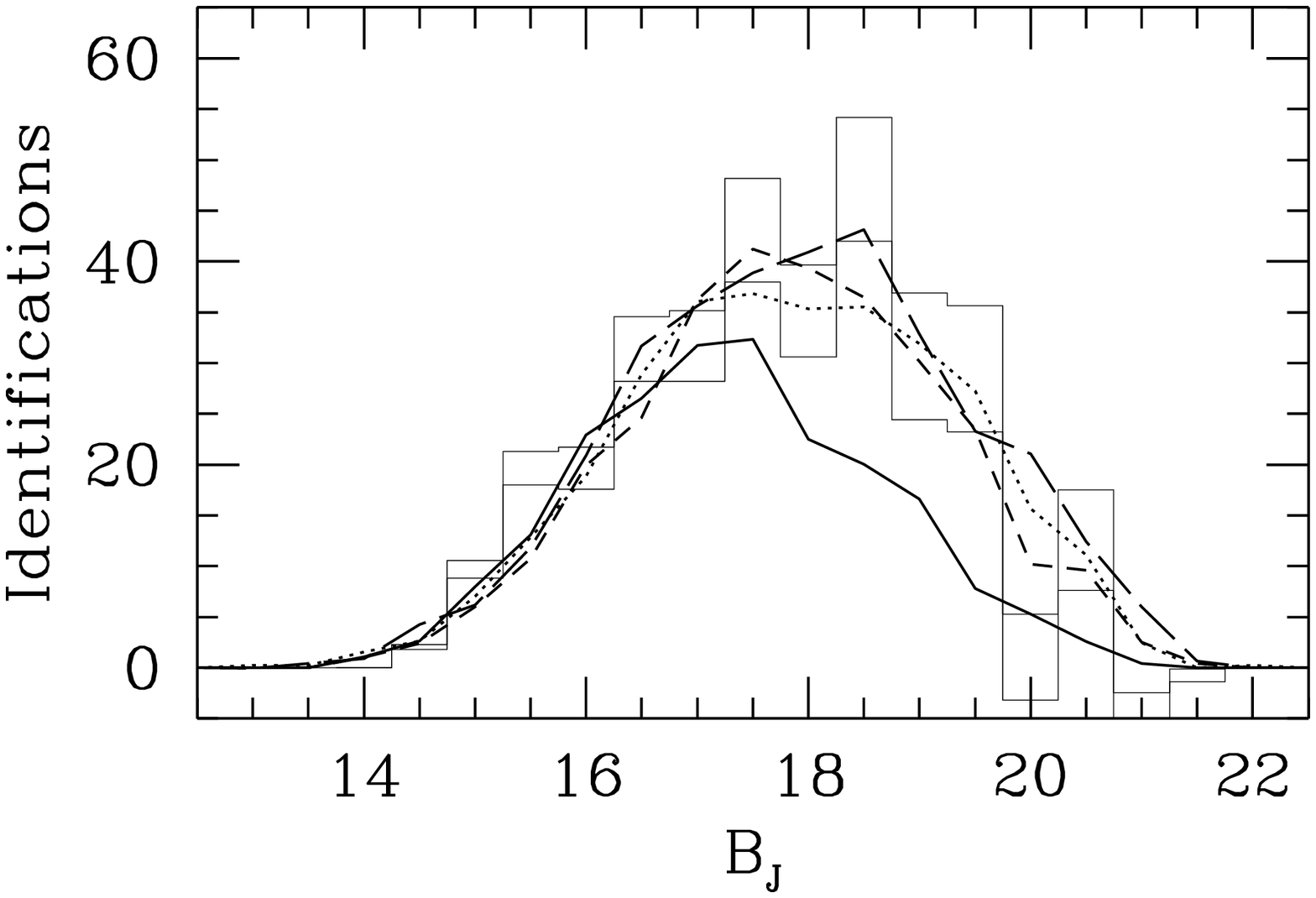,width=8.8cm}}
  \caption[]{
            Magnitude distribution of the optical counterparts to IRAS sources in the photographic $\Bj$
            band in two infrared flux domains. Up: $0.2 <f_{\nu}(60\mu{\rm m})< 1$~Jy, down:
            $0.11 <f_{\nu}(60\mu{\rm m})< 0.2$~Jy.
            The two histograms are the lower and higher limits to the observations induced by the
            correction of spurious identifications (see text), but do not include Poissonian uncertainties.
            The lines are the models as in
            Fig. \ref{fig:rawcounts}. Note that the optical catalogue is expected to become incomplete for $\Bj>20$
            and $\Bj\le 16$.
  \label{fig:magdis}
            }
   \end{figure}

  We now turn to comparing the magnitude distribution with model predictions. Once again, we have to
  subtract the contribution from fortuitous detections. One can decompose the total observed number of
  identifications per magnitude bin at magnitude $m$, $n(m)$, as
  \begin{equation}
  \label{eq:nm}
  n(m) = (1-p_f)n_r(m) + (1-p_r)n_f(m)+ n_c(m),
  \end{equation}
  where $n_r(m)$ is the number of real counterparts and $n_f(m)$ the number of fortuitous optical galaxies
  in the same magnitude interval. $n_c(m)$ represents the contribution from IRAS sources for which
  both real and fortuitous identifications can occur (with probability $p_rp_f$). We don't know $n_c(m)$,
  but we can safely bracket it between the two extreme cases:
  \begin{equation}
  0 \le n_c(m)\le p_fn_r(m)+p_rn_f(m).
  \end{equation}
  The lower bound corresponds to the case where the magnitude bin is totally depleted by the presence
  of the two ``competitors'', and the higher bound corresponds to the case where there is no depletion at all. The true value
  is quite probably intermediate. Inserting $n_c(m)$ in (\ref{eq:nm}) we get
  \begin{equation}
  n(m) - n_f(m) \le n_r(m) \le \frac{n(m) - \left( \frac{1-p}{1-p_f} \right)n_f }{1-p_f}.
  \end{equation}
  The magnitude distribution of IRAS sources with ${\cal L} > 5$ is
  plotted in Fig. \ref{fig:magdis}, together with model predictions, for two 60\mic\ flux domains.
  As expected from the evolution models, excess sources are found on the dim side of the histogram.
  A very good agreement is found for $3<Q<4$, but like the identification rate statistics, they do not put
  further constraints to the exact amount of evolution.

\section{Summary and comparison with other works}
\label{chap:comp}
  Although they cannot be considered as really independent one from each other, the results of the
  four statistical tests described above (background fluctuations, number counts, identification rate
  and magnitude distribution) do provide a picture remarkably consistent with the phenomenological
  $L^*\propto (1+z)^Q$ evolution model, with $3<Q<4$. The most stringent test is given by the number counts
  with $Q=3.2\pm 0.2 \pm 0.3$ down to 0.15 Jy, with some indication of a stronger evolution below 0.2 Jy.
  However, the fluctuation analysis indicates that $Q$ cannot increase much beyond $Q=4$ below 0.1 Jy.
  As indicated earlier, a pure density evolution yields a similarly good fit to the data, with a
  comoving density evolving as $\propto (1+z)^{6.0\> \pm 0.5\pm 0.7}$  down to 0.15 Jy.

\begin{figure}[htbp]
  \centerline{\psfig{figure=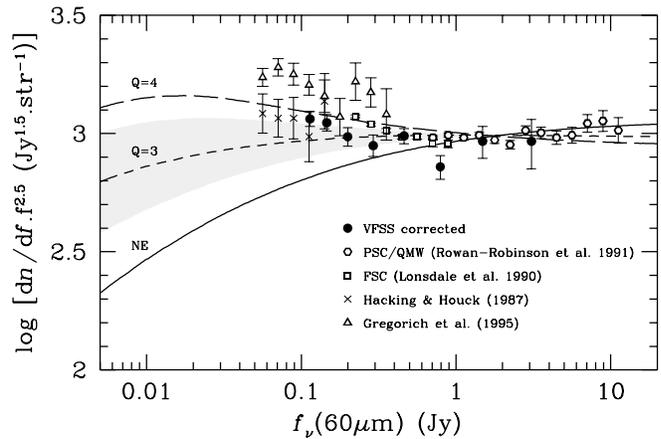,width=8.8cm}}
  \caption[]{
            Corrected VFSS number counts compared with previous studies and simple no-evolution (NE) and
            $L^*\propto (1+z)^Q$ models at 60\mic.
            The correction was done by subtracting to each bin ($f_{\nu}$(60\mic)$< 0.3$~Jy) the bias
            measured in Monte-Carlo simulations, with the $Q=3.5$ model. This bias depends weakly on
            the exact value of $Q$; for $3 \le Q \le 4$, the typical uncertainty is $\approx5$\%, and has been included
            in the error bars. Note that the other counts have not been corrected for the Eddington bias, and might be
            overestimated in their faintest bins. The grey area indicates approximately the $1\sigma$ confidence domain for
            number counts in the context of a $L^*\propto (1+z)^Q$ pure luminosity evolution model, drawn from the
            background fluctuation analysis (\S\ref{chap:fluctu}).
  \label{fig:corcounts}
            }
   \end{figure}

  These values are in good agreement with the value found on shallower surveys by Saunders \etal (\cite{saunders:ala})
  who found  $Q=3.1 \pm 1.0$ --- although according to some authors, e.g. Fisher \etal \cite{fisher:al} or
  Oliver \etal \cite{oliver:alb}, this number might largely be affected by biases ---,
  and Oliver \etal (\cite{oliver:alb}) with $Q = 3.3 \pm 0.8$. 
  However they do {\em a priori} rule out the extreme evolution discovered by Gregorich \etal
  (\cite{gregorich:al}) in their IRAS number counts. In particular, their high density of faint sources is
  incompatible with our background fluctuation analysis. A check done on the Gregorich \etal
  (\cite{gregorich:al}) fields reveals that they contain cirrus, which might produce
  a large fraction of false detections, and should also contribute significantly to the total background noise and
  its associated Eddington bias. Strong cirrus emission also affects the Hacking \& Houck
  (\cite{hacking:houck}) deep field, but the dispersion in their scanning angles smoothes considerably
  edge-effects on cirrus structures. This might explain why their counts are in better agreement with ours.
  Our counts and those from previous studies are summarized in Fig. \ref{fig:corcounts}.

\section{Discussion}
\label{chap:discu}
\subsection{The nature of the faint source excess}
  Our statistics probes mainly the evolution of infrared luminous galaxies ($L_{60}>L^*$).
  The good fits given to the data by the $L^*\propto (1+z)^{3.2}$ model, and the large excess of faint optical
  counterparts indicate that luminous IRAS galaxies should be $\approx\,$5-7~times more numerous at $z\approx 0.3$
  than at the present time.
  What are these galaxies? According to our Monte-Carlo model, the optical counterparts with $20<\Bj<21$
  should have an average 60\mic\ 
  luminosity $L_{60} \approx 5.10^{11} h^{-2} L_{\odot}$ and should be at a redshift $z\approx 0.3$,
  in agreement with their
  rather red colours $\Bj-\Rf \approx 1.6$, if their optical continuum is close to that of local Sbc's.
  These are therefore potential UltraLuminous InfraRed (ULIRGs) candidates.
  This picture is consistent with the redshift distribution of the optically faint ULIRGs surveyed by
  Clements \etal (\cite{clements:ala}).
  Objects with brighter optical counterparts
  ($\Bj<20$) have naturally a lower FIR excess, but most are still {\em redder} than the average population in
  the same magnitude domain. They are very probably similar to the population of
  starburst spirals with $\approx M^*$ optical luminosities identified in the sub-milliJansky radio
  surveys\footnote{Incidentally, the evolution of sub-mJy radio sources seems compatible with that of IRAS galaxies
  (Rowan-Robinson \etal \cite{rowan:alc}).} (Benn \etal \cite{benn:al}, Windhorst \etal \cite{windhorst:al}), although
  at smaller redshifts ($z \approx\,$0.1-0.2).
  Hence it seems unlikely that a large fraction of the detected excess of faint 60\mic\ sources at the 100 mJy level might
  be caused by dust-enriched primeval galaxies, as suggested by Franceschini \etal (\cite{franceschini:alb}).

  Wang (\cite{wang1}, \cite{wang2}) proposed a pure FIR-luminosity evolution scenario based on the evolution of dust
  content in normal galaxies. Although some variants of this model might be able to predict a significant increase of FIR luminosity
  in galaxies over the few past Gyr, they rely on the assumption that emitters are
  optically thin at blue/UV wavelengths. This is certainly not the case for most of FIR-luminous galaxies
  (the sole population of galaxies that can be detected by IRAS at significant redshifts).
  As these objects already have $L_{\rm FIR} > L_{\B}$ , adding more dust should not be very effective in
  increasing their FIR luminosities (e.g. Belfort \etal \cite{belfort:al}).
  Therefore a moderate change in the dust content of galaxies since $z\approx 0.3$ should not be a major
  contributor to the evolution observed yet for IRAS galaxies.

  The evolution rate of luminous IRAS galaxies is in remarkable agreement with those of QSOs
  (e.g. Boyle \etal \cite{boyle:al}) or X-ray selected Active Galactic Nuclei (AGN)
  (e.g. Maccacaro \etal \cite{maccacaro:al}), which brings up the link between AGN and
  ULIRGs, as most of the latter may harbour AGNs (e.g. Sanders \etal \cite{sanders:alb}).
  An important question is therefore: does the evolution concern the global FIR-luminous galaxy
  population, or is it limited to ``monsters'' containing active nuclei? The magnitude distribution in
  our faintest 60\mic\ flux bin (Fig. \ref{fig:magdis}) seems to indicate that most of the sources found
  in excess are not ULIRGs,
  and consequently, according to local statistics (Leech \etal \cite{leech:alb}), are less likely to contain an AGN.
  A stronger argument against a sample dominated by AGNs may come from spectroscopic identifications of sub-milliJansky
  radio-sources (Benn \etal \cite{benn:al}), which, as we saw, can be compared to our sample, and has proven to contain a
  minority of AGN-like spectra. Extreme caution should however be made in interpreting these spectra, as optical signatures
  of a central engine might be hidden by the large amounts of dust present in the nuclear regions of such objects (see
  e.g. Dennefeld \cite{dennefeld}).

\subsection{The link with optical evolution}
  Deep number counts conducted at visible wavelengths have revealed
  a large excess of faint galaxies over the no-evolution predictions
  from $q_0>0$ cosmological models (Tyson \cite{tyson}, Colless \etal \cite{colless:al}, Lilly \etal \cite{lilly:al},
  Metcalfe \etal \cite{metcalfe:ala}, \cite{metcalfe:alb}, Smail \etal \cite{smail:al}),
  as well as a blueing of the global population with fainter magnitudes.
  The $z<1$ redshift distributions obtained from spectroscopic surveys
  are however apparently in good agreement with no-evolution models,
  which suggests a diminution or a fading with time of low and moderate luminosity members
  of the blue galaxy population\footnote{For simplicity, we shall refer to these objects as ``faint blue galaxies''.}
  since $z\approx 0.5$ (Broadhurst \etal \cite{broadhurst:ala}, Lilly \cite{lilly},
  Ellis \etal \cite{ellis:al}, Driver \etal \cite{driver:ala}).
  Various scenarios to explain this observational phenomenon have been proposed, including a delayed
  formation epoch of dwarf galaxies (Babul \& Rees \cite{babul:rees}, Babul \& Ferguson \cite{babul:ferguson}),
  decimation through intense merging (Guiderdoni \& Rocca-Volmerange \cite{guiderdoni:rocca},
  Broadhurst \etal \cite{broadhurst:alb}), or more prosaically, biases in the determination of the
  local luminosity function (Davies \cite{davies}, McGaugh \cite{mcgaugh}, Gronwall \& Koo \cite{gronwall:koo}).

  Now, strong evolution among the spiral population, apparently seen at FIR and radio wavelengths,
  is a tempting argument for hypotheses advocating recent evolution of the optical LF.
  Evolution in FIR is detected in redshift surveys of IRAS samples (Saunders \etal \cite{saunders:ala},
  Oliver \etal \cite{oliver:alb}) and could be explained, for example, through an increasing
  rate with $z$ of galaxy-galaxy interactions, which are known to be efficient triggers of starburst
  activity (e.g. Larson \& Tinsley \cite{larson:tinsley}), or AGNs (Sanders \etal \cite{sanders:ala}).
  There are indeed both theoretical (Toomre \cite{toomre}, Carlberg \cite{carlberg}) and observational
  (Zepf \& Koo \cite{zepf:koo}, Burkey \etal \cite{burkey:al}, Carlberg \etal \cite{carlberg:al})
  evidences for such a rapid increase with redshift
  of the merging rate, at least to $z \approx 0.7$.
  A few attempts have been made to link a strong,
  starburst-driven, evolution of the FIR luminosity function to that of the optical
  (Lonsdale \& Harmon \cite{lonsdale:harmon}, Lonsdale \& Chokshi \cite{lonsdale:chokshi},
  Pearson \& Rowan-Robinson \cite{pearson:rowan}).
  They involved simple proportionality between the optical
  and FIR evolution (either in density or luminosity) and therefore predicted a large increase in the
  comoving space density
  of optically bright ($M < M^*$) galaxies with redshift, which is unfortunately not seen at $z<0.5$.
  The same problem arises
  with stellar population synthesis models applied to starburst-driven evolution (e.g. Carlberg \& Charlot
  \cite{carlberg:charlot}).

  But this apparent contradiction could be alleviated by remarking that the 
  possibly large increase of blue continuum emission
  resulting from starburst activity is often largely {\em hidden} at blue/UV wavelengths in FIR luminous
  galaxies. It is a fact that the complex role of dust, and in particular its geometry with respect to the stars,
  has been ignored in many models trying to reproduce number counts with starburst-evolution.
  In massive spirals, star-formation induced by merging is generally concentrated in the central
  regions of the galaxy (e.g. Hummel \cite{hummel}, Condon \etal \cite{condon:al}, Bushouse \cite{bushouse}).
  Such a feature is likely to originate from radial inflow of disk gas triggered by the interaction, and is
  well reproduced in numerical simulations (e.g. Hernquist \cite{hernquist}, Minhos \& Hernquist \cite{minhos:hernquist}).
  The associated circumnuclear
  starburst (which may surround an AGN) is usually heavily reddened in the optical, and does not increase so much
  the total blue/UV light output of the galaxy: ULIRGs are rarely ``ultraluminous'' at visible wavelengths (as
  reflected by relation \ref{eq:bilf}).
  On the contrary, starbursts in field dwarf galaxies, dominated by
  gas-rich, dust-deficient late types (e.g. Van den Bergh \& Pierce \cite{vdbergh:pierce}, Wang \& Heckman
  \cite{wang:heckman}) typically show up in optically prominent giant H~II regions (e.g. Huchra \cite{huchra}),
  and are often found to dominate the total blue/UV emission. Hence, statistically, starbursts
  observed in both types of galaxies possess very different observable signatures. Therefore the hypothesis that
  starburst galaxies which dominate IRAS faint number counts are the same as those that make the excess seen
  in the optical at intermediate redshifts (Pearson \& Rowan-Robinson \cite{pearson:rowan}) is not
  convincing.

  There is nevertheless evidence from optical redshift surveys that the galaxies with $M \ge M^*$ observed in
  excess at $z\la 0.5$ are actively star-forming galaxies (Ellis \cite{ellis:al}, and references therein).
  Recent studies have estimated their evolution in terms of global brightening at blue wavelengths
  between now and $z\approx 0.3-5$ to $\approx 1-1.5$~mag (Driver \etal \cite{driver:alb}, Rix \etal \cite{rix:al}), similar
  to what we inferred for IRAS galaxies in FIR.
  Is the evolution of FIR-luminous galaxies closely related to that of faint blue galaxies?
  In the hypothesis of a recent evolution dominated by starburst processes in both populations, this question is essentially
  linked to the existence of a common triggering mechanism, in which case we may consider IRAS galaxies as massive,
  dusty versions of the faint blue galaxies.
  Although interaction/merging seems to be the rule among distant IRAS galaxies, at least for the most luminous
  members (Clements \etal \cite{clements:alb},\cite{clements:alc}), the situation is much less clear for faint blue galaxies.
  Deep {\em Hubble Space Telescope} images present visual evidence for a high proportion of
  interacting/merging objects, increasing steeply with magnitude (Driver \etal \cite{driver:ala}, van~den~Bergh \etal
  \cite{vdbergh:al}). Nevertheless, many of these sources may well be luminous galaxies at much higher redshifts,
  or in some cases, knots of star-formation observed in single objects. Besides, optical observations of nearby dwarf galaxies
  undergoing starbursts reveal that a large fraction are apparently isolated objects (Campos-Aguilar \& Moles \cite{campos:moles},
  Telles \& Terlevich \cite{telles:terlevich}), which argues against interactions as a unique triggering mechanism of
  starburst activity at recent epochs. Still, one cannot exclude the hypothesis of interactions or collisions with HI clouds,
  supported by 21~cm observations of the environment of nearby HII galaxies (Taylor \etal \cite{taylor:al}).

  Another test of the interaction scenario might be provided by studying the spatial (or angular) two-point correlation
  function. One could expect the distribution of sources triggered by interactions to exhibit some differences with respect
  to the global (late-type) density field. In fact both IRAS and blue galaxies happen to
  have clustering properties barely distinguishable, within measurement errors, from their common parent
  population of late-type galaxies (e.g. Infante \& Pritchet \cite{infante:pritchet}, Mann \etal \cite{mann:al},
  Oliver \etal \cite{oliver:alc}), although this may not be true at very small scales (Carlberg \etal \cite{carlberg:al},
  Infante \etal \cite{infante:al}).

  We shall therefore conclude this discussion by pointing out that a simple and appealing scenario, in which the
  strong, recent evolution among late-type galaxies at moderate redshifts is related 
  to an increase of the merging/interaction rate with lookback time,
  appears qualitatively compatible with the observations. In this scenario, 
  FIR-luminous galaxies would then represent the
  massive, dusty counterparts of faint starbursting blue galaxies.
  More data are clearly necessary to test this hypothesis on the observational side, under others to follow
  the evolution of the intermediate luminosity population in the FIR. This should be 
  possible with the ISO ({\em Infrared Space Observatory})
  satellite. It would also allow to assess more accurately the evidence 
  for strong evolution at low redshifts in the FIR, deduced so far only from IRAS measurements.

\section{Conclusions}

  We have analysed a deep homogeneous 60\mic\ subsample of galaxies, at the
  lowest flux limit reached by the IRAS Faint Source
  Survey, yielding the following results:

  1. Detection and measurement biases were assessed through Monte-Carlo 
  simulations, which prove to reproduce satisfactorily
  most observed features of detected IRAS sources at 60\mic. In particular,
  the rather high proportion of unidentified sources
  in the FSC ($\ge 10$\%) can totally be accounted for by deflections caused by
  neighbouring IR sources.

  2.  Background fluctuation analysis and number counts provide evidence for
  strong evolution in FIR luminosity
  $\propto (1+z)^{3.2\> \pm 0.2\pm 0.3}$ or density $\propto (1+z)^{6.0\> \pm 0.5\pm 0.7}$,
  in agreement with previous studies.

  3. Sources in excess are genuine and are generally associated with faint,
  relatively red, optical counterparts which
  we interpret as being $M^*$ massive starbursting galaxies at redshifts $\ga 0.1$.

  4. Photometric properties of distant IRAS galaxies suggest that the
  advocated starbursts (or AGNs) are considerably
  hidden by dust at UV/visible wavelengths, which would explain why no large
  number of related optically luminous objects
  is showing up in optical surveys at $z\le 0.5$.

  5. Faint IRAS galaxies in excess are unlikely to be the ones
  making the excess in number counts at blue wavelengths and
  $z\le 0.5$, but this does not exclude that both populations follow the same
  evolution mechanism (i.e. interaction-induced starbursts).

\begin{acknowledgements}
The authors wish to thank A. Fruscione for her contribution during an earlier phase of the project, and C. Lidman
for helpful comments on the manuscript.
One of us (EB) acknowledges an ESO studentship while part of this work was completed.
\end{acknowledgements}

\end{document}